\documentclass[12pt,preprint]{aastex}

\shorttitle{BUDDA: Bulge/Disk Decomposition Analysis}
\shortauthors{de Souza, Gadotti, \& dos Anjos}

\begin{document}

\title{BUDDA: A New Two-Dimensional Bulge/Disk Decomposition Code for Detailed Structural Analysis
of Galaxies\footnote{Based on observations made at the Pico dos Dias Observatory (PDO/LNA -- CNPq),
Brazil}}
\author{R. E. de Souza, D. A. Gadotti, and S. dos Anjos}
\affil{Departamento de Astronomia, Instituto de Astronomia, Geof\'{\i}sica e Ci\^encias
Atmosf\'ericas, \\ Universidade de S\~ao Paulo \\ Rua do Mat\~ao, 1226 -- Cid. Univers.
CEP 05508-900, S\~ao Paulo -- SP, Brasil; ronaldo@astro.iag.usp.br, dimitri@astro.iag.usp.br,
sandra@astro.iag.usp.br}

\begin{abstract}
We present {\sc budda} (BUlge/Disk Decomposition Analysis), a new code devoted to perform a two-dimensional
bulge/disk decomposition directly from the images of galaxies. The bulge component is fitted with a generalized
S\'ersic profile whereas disks have an exponential profile. No other components are
included. Bars and other sub-structures, like lenses, rings, inner bars and inner disks,
are studied with the residual images obtained through the subtraction of bulges
and disks from the original images.
This means that a detailed structural analysis of galaxies may be performed with a small number
of parameters, and sub-structures may be directly studied with no a priori assumptions.
As has been already shown by several studies, two-dimensional fitting is much more reliable than
one-dimensional profile fitting. Moreover, our code has been thoroughly tested with artificial data and we
demonstrate it to be an accurate tool for determining structural parameters of galaxies. We also show that
our code is useful in various kinds of studies, including galaxies of, e.g., different morphological types,
and inclinations, which also may be observed at different spatial resolutions. Thus, the code has a broader
range of potential applications than most of the previous codes, which are developed to tackle specific problems.
To illustrate its usefulness, we present the results obtained with a sample of 51 mostly early--type galaxies
(but covering the whole Hubble sequence).
These results show some of the applications in which the code may be used: the determination of parameters
for Fundamental Plane and structural studies, quantitative morphological classification of galaxies and
identification and study of hidden sub-structures. We have determined the structural parameters of the galaxies
in our sample and found many examples of hidden inner disks in
ellipticals, secondary bars, nuclear rings and dust lanes in lenticulars and spirals, and also wrong
morphological classification cases. We now make {\sc budda} generally available to the astronomical community.
\end{abstract}


\section{Introduction}
It is widely assumed that the light distribution in galaxies may be described by one or more components,
which might be dynamically and/or chemically distinct. While most of ellipticals (E) may
be photometrically described by only one major spheroidal component, lenticular (S0)
and spiral (S) galaxies in general need at
least two components: a spheroidal bulge and a flattened disk.
However, in almost every object we can
trace the presence of other relevant components, such as primary bars, secondary bars (in spirals
and S0s), inner disks, lenses and rings, none of which have been as carefully studied as
the bulges and disks.
Although there is an unclear and poorly
understood interplay between these components, which is not generally taken into account in most structural analyses,
bulge/disk decomposition has been a major motif
behind a number of important contributions to our
understanding on galaxy formation and evolution. Examples include the Tully-Fisher relation \citep{tul77}, the
Fundamental Plane \citep[hereafter FP]{fab87,dre87,djo87,ben92}, the correlation
between the scale lengths of bulges
and disks \citep{cou96,dej96b,anj04} and the evolution of galaxy morphologies \citep{mar98}. See also
\citet{pen02} and \citet{gra02a,gra02b} for further and more recent examples.

Two important issues are related to the subject of
bulge/disk decomposition: the fitting method and the mathematical
functions used to describe each component. Generally speaking, there are two fitting methods, the
one-dimensional (1D) and the two-dimensional (2D). In the 1D method, an azimuthally averaged surface
brightness profile of the galaxy under study, or the major or minor axis profiles,
is fitted by one or more components. This method has the
advantage of being simple and fast, and works in low S/N conditions. However, it has two major problems.
Firstly, the procedure to obtain the galaxy profile may vary, and each procedure has its own drawbacks.
In isophote fitting \citep{jed87}, which is commonly used, changes in ellipticity and position angle
at consecutive isophotes turn the profile poorly defined, since it is extracted not along a straight line but an
arc. This is especially true when a bar or a strong spiral arm is present.
Alternatively, one may extract the profile along the galaxy major axis, which may be itself poorly defined,
but in this case much of the information in the image is thrown away \citep{bag98}. Moreover, minor axis profiles
are also essential in this case since the bulge dominates the profile along this axis. Secondly, 1D
fitting averages out in a complicate way non-axisymmetric components, like bars.
It is possible, though, to obtain reasonable results with 1D fitting, as shown by \citet{ken89}, which
study the SB0 galaxy NGC 936 by means of 1D decomposition but also incorporating further intricate steps
\citep[see also][]{ken86}.
In 2D fitting, information in the whole image is
used to build a model for each component and thus differences in the projected ellipticities and position
angles become an advantage, since they help to constrain the parameters of the different components.
Moreover, non-axisymmetric components are now taken into account. There are several examples in the
literature of studies showing that the 2D method is much more reliable than the 1D
\citep{dej96a}, retrieving more accurate structural parameters.

While there are several examples in the literature of studies which use 1D fitting \citep{bag98}, only
in the last decade 2D fitting algorithms become more popular, mainly because of the higher
computational speed of today's machines. As examples, we
may point out the algorithms developed by \citet{dej96a},
by \citet{sim98}, and more recently by \citet{pen02} and \citet{tru01}. Most of the
available algorithms were developed to tackle specific problems, like, e.g., face--on late--type spirals
\citep{dej96a}, distant galaxies \citep{sim98}, or nearby galaxies observed with very high spatial resolution
\citep{pen02}. On the other hand, it is a common practice to include several components other than the
spheroid and the disk in an attempt to get better fits. In this way, one can model bars and, in the case of
\citet{pen02}, also nuclear point sources, inner disks and inner bars, and lenses. In the Peng et al.
algorithm 40 free parameters are used, and in several cases a single component is fitted by various
different functions.

A caveat behind this philosophy of performing structural analysis is related to the fact that we do not
know how well a single general function could fit these sub-structures with a given accuracy. Therefore
these fits might be subject to an unknown systematic effect.
For instance, as some bars show an exponential profile, especially in late--type galaxies, and others
have a flat light distribution, especially in early--type galaxies \citep{elm96}, with a steep outer drop,
it is difficult to foresee how a single general function could be used to fit
bars and give meaningful structural parameters. This
is also the case for lenses and other sub-structures. In our code,
we explore the alternative route of fitting the two
major components, representing the disk and the bulge contributions, and to
perform a detailed study of sub-structures in the residual images obtained
after the subtraction of these major components. In principle, this procedure
could allow us to study the minor structures and assess the possibility to
describe them as families of functions.
It should be noted, however, that residual images can also be obtained with 1D fits if one
can derive reasonable estimates of the apparent axis ratios of the bulge and disk components.
For instance, \citet{but92} show a residual image of NGC 4622 based on 1D decomposition
expanded to 2D using assumptions about its components
projected shapes. Also, 2D bulge/disk decomposition
is not the only way to bring out hidden features. Using Fourier decomposition \citep[see, e.g.,][]{con88},
\citet{but92} were able to bring out remarkably well the amazing counter-winding spiral
arms in NGC 4622. To a lesser extent, a similar analysis can be performed via unsharp
masking and color maps \citep[see, e.g.,][for a recent application]{erw02}.

Therefore, following these guidelines,
we have developed a new and general purpose algorithm, {\sc budda}\footnote{See \anchor{
http://www.astro.iag.usp.br/~dimitri/budda.html}{\url{
http://www.astro.iag.usp.br/$\sim$dimitri/budda.html}}
where {\sc budda} is available to the astronomical community.} -- which stands for BUlge/Disk
Decomposition Analysis -- devoted to perform a detailed 2D
structural analysis in galaxy images. The code allows us to perform a direct study on sub-structures in galaxies by
means of accurate residual images, besides determining structural parameters of the main components,
namely the bulge and the disk. In Sect. 2 we present the details of how the algorithm works.
In Sect. 3 we present a
thorough analysis of the reliability of the algorithm using a set of artificial data images of galaxies having known
structural parameters. In Sect. 4 we apply our algorithm to a
sample of 51 galaxies and show the results. Finally, in Sect. 5 we finish with a summary and a
brief discussion of our results and the potential applications of our code.

\section{The Algorithm}
From the point of view of our algorithm, a galaxy is simply an image
composed by just two major components: the bulge and the disk. In
order to extract the properties of these two components we need to describe
them in a generic way and predict their expected contribution at each pixel
of the observed image. For the disk we have adopted the exponential profile \citep{pat40,fre70}:

\begin{equation}
{I_{d}(r)} = {I_{0d} e^{-r/h}},
\end{equation}

\noindent where $h$ is the disk scale length, $I_{0d}$ represents the
central brightness and $r$ is the semi major axis of the elliptical isophotes
having ellipticity $e_d$. Both components, the bulge and the disk, share a
common center located at $x_0,y_0$, but the ellipticity ($e_d$, $e_b$) and
position angle ($PA_d$, $PA_b$) are not necessarily equal. In principle, we
expect that the information on the shape parameters of the disk is mainly
sampled in the outermost isophotes of the galaxy, while the bulge is more
influential in the innermost central region. Therefore, by tracking the subtle
pixel variations due to changes in the isophotal shape going from the center to
the external region the program is able to model both the bulge and disk
isophotal shapes.

Bulges are usually described in the literature by the $r^{1/4}$
profile or de Vaucouleurs' function. Several examples of elliptical galaxies
are known to accurately obey the de Vaucouleurs' law over as much as 10
intervals of magnitude in surface brightness. On the other hand, there are also
several objects, including bulges, where this relation does not provide an
accurate description. For these reasons we rather prefer the use of the so
called S\'ersic profile, which is a more generic mathematical expression that
includes the de Vaucouleurs profile as a special case. Moreover, there is
strong evidence that the S\'ersic profile is more efficient in the empirical
description of isophotal properties of early type galaxies \citep{cao93}.
Therefore, we adopt the solution of describing the brightness profile of the
bulge by using the relation

\begin{equation}
I_b(r) = I_{0b}10^{-b_n[(r/r_{eb})^{1/n}]},
\end{equation}

\noindent where $r_{eb}$ is the effective radius, $I_{0b}$ is the central
brightness and $n$ is an index controlling the shape of the brightness profile.
For $n=4$ we recover the de Vaucouleurs profile, while for $n=1$ we have the
exponential profile. Therefore the index $n$ is an important quantity
describing the radial variations of the brightness profile. The numerical
constant $b_n$ is defined in such a way that the the effective radius,
$r_{eb}$, contains half of the total bulge luminosity and its value is
approximately given by $b_n \simeq 0.868 n + 0.142$. At $r=r_{eb}$,
$I_b=I_{eb}$, the bulge effective surface brightness. The coordinate $r$
indicates the semi-major axis of the bulge elliptical isophotes centered on
$x_0,y_0$, having an ellipticity $e_b$ and a position angle $PA_b$.

The central brightness profile is known to be heavily affected by the
atmospheric seeing. Therefore we have included in the code a circular gaussian smearing of
the brightness profile controlled by a parameter associated with the seeing
radius ($a_s$). This is similar to what was already done by \citet{dev79} and \citet{cap83}.
Basically it affects only the very central region where the convolution
with the original profile is able to flatten the central brightness distribution.
But its inclusion is quite important because otherwise we would get very
discrepant models in the central region, changing the optimum minimum $\chi^2$
solution. We have also included in the code a supplementary correction of the
sky level. In fact, all images were sky corrected before applying the code. But
the fact is that even in a careful analysis the sky subtraction could be quite
important mainly in the outer region where the disk and bulge brightness
profiles come close to zero. An oversubtraction of the sky contribution
could easily mimic a truncation of the external region of the disk, while an
undersubtraction can be misinterpreted as an extended disk. Therefore, we have
included this correction term ($\Delta_{sky}$) in order to verify if the sky correction was done
properly or not. Whenever this correction was higher than a few percent of the
sky level we have reanalized the sky subtraction procedure to verify the
reasons behind such discrepancies. In some cases, the sky subtraction might be
difficult to determine due to the projection of nearby stars,
companion galaxies, or simply due to the large size of the galaxy in the frame
that precludes a good sampling of sky regions. These worse cases, where we could
not improve the sky subtraction procedure, have been removed from our analysis.
Note that we have done a careful sky subtraction in the galaxy images studied
in this paper (see Sect. 4) and that these images of our sample of galaxies
generally allow for a proper determination of the sky background. Thus, the cases where
the $\Delta_{sky}$ term suggests that the sky subtraction may have failed were
indeed very rare in this study.

It is also possible to fit edge--on galaxies. In this case, the disk ellipticity is not
anymore a useful parameter and it is discarded by the code. Instead, it uses
a vertical light distribution for the disk which follows \citet{van81}, i.e.,
$\propto \rm{sech}^2 (z/z_0)$, where $z_0$ is the isothermal scaleheight. Moreover, the
ellipses describing the bulge and the disk components do not need to be necessarily
perfect: boxy and disky isophotes are allowed, and the code looks for the best
description. These are described by generalized ellipses and the ellipse index parameter
$\gamma$ controls their shape: for $\gamma=1$ the ellipses are perfect, while for $\gamma>1$
the ellipses are boxy; disky ellipses have $\gamma<1$.

The user also has the option of considering an inner truncation in the disk and an outer truncation
in the bulge. Note, however, that in this study we have not restricted the contributions of both
bulge and disk in any region of the galaxies whatsoever. We have added these truncations
as user specifications in the code to allow for the freedom of studying, for instance, disks
with type II profiles \citep{fre70}. Any user should be very careful, though, in using these
parameters, by making a priori assumptions that a given area
of the galaxy is dominated by the bulge or the disk, as was often done with the 1D fits.
These truncations may also work as a way to exclude certain regions from the 2D fit.
We stress again that this has not been done in this paper.
For instance, this is not what we have done to deal with galaxies with bars or other conspicuous
components. To model such cases, we rely on the code's ability to distinguish bars and
spiral arms from the bulge and the disk by means of differences in the position angle, ellipticity and
luminosity profile. Unless these three constraints are all equal for, say, the bulge and the bar, the
code is able to separate these components, as is revealed by the analysis of barred galaxies
(see Sect. 4), in which the combined luminosity profile of the bulge and disk models falls below
the observed profile for the galaxy only in the bar region.

Altogether the code needs to determine a total number of 11 main
parameters: $x_0$, $y_0$, $I_{0d}$, $h$, $e_d$, $PA_d$, $I_{0b}$,
$r_{eb}$, $e_b$, $PA_b$, and $n$. The others, $a_s$, $\Delta_{sky}$, the two
$\gamma$ indices (one for bulge and one for disk), and the two truncation radii (the outer one for bulge
and the inner one for disk) may be set constant in general. An initial guess for these
parameters, together with an initial allowed range of variation, are provided
by the user in an input file read by the program. If the range of a given
parameter is initialized as zero the code understands that it is a fixed
parameter that should be kept constant during the iteration process. This is a
useful way of dealing with parameters that are difficult to be determined even
by using the whole image information. For instance, if we have an almost
circular component the position angle becomes badly constrained and difficult
to be determined. In these cases we have opted to have an initial run of the
code, keeping these parameters fixed, and then a second run was made leaving
them unconstrained. The program also expects to receive from the user
information about the noise amplitude of the image being analyzed. This is a
crucial piece of information since the code will decide the best solution by calculating
the $\chi^2$ of the overall fit of the model. This information comes in two ways:
the flat noise level and the Poisson noise level. The former is a number in counts
that represents the noise in flat regions of the image, which is dominated by contributions
from the noise in the sky level and from residuals due to imperfections in the
flatfield normalization. The flat noise level can be estimated by the user or by
the code itself; the minimum valid count for the code will be the square root of the flat noise level.
The Poisson noise level is kept to 1 if the user wants to assume that the noise in the galaxy image is
purely Poissonic. The code may also verify if this is a correct assumption and change this level
accordingly. To avoid complicated regions in the image the
user can flag those regions dominated by stars, defects or other companion
objects that could spoil the solution. This may be done with the {\sc iraf}\footnote{{\sc iraf}
is distributed by the National Optical Astronomy Observatories,
which are operated by the Association of Universities for Research
in Astronomy, Inc., under cooperative agreement with the National
Science Foundation.} task {\sc imedit}. Note that while only bright stars may interfere
in the code's determination of the bulge and disk parameters, the fainter ones still contribute
to the calculation of the $\chi^2$. Thus it is very important that the galaxy image be as free
as possible from this kind of perturbation. Moreover, the maximum radius of the
galaxy can also the informed so that the code doesn't try to find information
outside the region where the object is present. At the end of this initial step
the code selects all valid pixels that will be used during the fitting procedure.

The fits are done in raw intensity units and all pixels have the same weight.
To obtain the optimum solution we have adopted the multidimensional
downhill simplex method \citep{pre89}. The iteration scheme works simply by
testing all the possible combinations of the parameters in each step and
choosing the one having the smallest $\chi^2$. In each step, the
multidimensional volume along the n-dimensional parameter space is preserved.
If a minimum is found along any given direction in the parameter space, the
program performs a parabolic interpolation to improve the estimate for the
next step. When a global minimum is found, the program does a final parabolic
interpolation and estimates the error along any given parameter using the {\em maximum}
$\chi^2$ variation allowed by the confidence level demanded by the user.
Thus, it must be noted that the errors at each parameter determined by the algorithm can
sometimes be overestimated. This is not related to the fact that the decomposition is a 2D one,
but only because the code uses the maximum $\chi^2$ variation allowed.
In the end, the code generates a file containing the expected image for each
component that could be added and compared with the original data.

The code is available in two executable task files: {\sc gmodel} and {\sc bmodel}. The former is
the one which determines the best bulge and disk models for the galaxy under study; the
latter builds synthetic images from the results of the former.

\section{Tests}
To verify the reliability of the {\sc budda} algorithm we have performed a long series of tests. Firstly, the code
was compiled on different machines (PCs and workstations) and under different operational systems (Linux and
Unix). As there are small differences between Linux and Unix Fortran compilers, we have made small changes
in the code so as to have proper versions for Linux and Unix. The different versions were tested and it was checked
that they give the same results in any machine when applied to the same galaxy, whether a real object from
our sample or a synthetic generated model.

To certify that the code reliably retrieves the right structural parameters of bulges and disks in galaxies, we
have applied it to 41 synthetic galaxies. These were built with the {\sc bmodel} executable within {\sc budda}
so that we know the galaxies' parameters a priori and can compare them with the results when we apply the
{\sc gmodel} {\sc budda} executable to the galaxies' images. The synthetic galaxies are made of a bulge and
a disk, with no other structural component such as a bar, since the main goal of the tests was to verify
the reliability in the values obtained for the structural parameters of bulges and disks. While it is true that
the complexity of real galaxies, which contain bars, spiral arms etc., may cause difficulties in the retrieving
of the correct parameters, we have argued above that the code is able to distinguish between the surface
brightness contributions from different components through constraints such as the ellipticity, position angle
and luminosity profile. Thus, bars and spirals are not considered by {\sc budda} and then appear
more conspicuously in the residual images. Moreover, when our code is applied on such more complex
galaxies, the determination of the errors in the estimates of the structural parameters is made
accordingly, since it takes into account the value of the $\chi^2$ for the fit (see Sect. 2).
The code was applied exactly in the same way as it is done for
real galaxies. The first guesses for the structural parameters were randomly chosen so that they can
initially be quite different from the original value used in
the simulation. No matter how far from the right value the first guess
for a parameter may be, the code gradually tends to converge to the right value, within the expected
error. This shows
how robust the results from the code are, even if one doesn't have a good estimate about the morphological
properties of the galaxy under consideration before the algorithm is applied.

We have done two kinds of tests. In the first one, a good spatial resolution image was simulated, similar
to most of the studies using ground-based telescopes without active or adaptive optics.
For these tests we have adopted a formal resolution of a two arcsec
seeing in a galaxy image with a diameter of around 2 arcmin. The results we present in the next section are
based on observations generally somewhat better than this. The tests show that this resolution is already
good enough to determine structural parameters with excellent reliability (see below). To check how {\sc budda}
works in worse cases (e.g., imaging of dwarf galaxies or galaxies at intermediate redshifts), we applied the same
tests with a resolution two times poorer than that.

It is worth noting that in these tests there was no user intervention, as should be the case with real galaxies,
when some actions are recommended to be taken in advance in order
to maximize the code efficiency in retrieving accurate
data. For instance, in real galaxies
{\sc gmodel} should be applied several times, choosing different sets of initial values, to search for
the best $\chi^2$ result. That precaution avoids the adoption of a solution
that could be located in a local minimum. In some cases, one may initially fix some geometric parameters
by other means; for instance, we could guess the initial position angle
or ellipticity with the {\sc ellipse} task from
{\sc  iraf}. Then, after a first solution is found, one may leave these parameters free to vary
in a second run, using the initial values as those of the solution of the
first run. This is the recommended procedure,
for instance, when the ellipticity is too small (around 0.1) and
the position angle is difficult to determine. However, none of these actions was done in the tests, so that their results are actually a
lower limit for the code's accuracy. Poissonic noise was added in some of the synthetic images but we found
it to be a minor source of error, as long as the S/N ratio is reasonably high (i.e., around 50
in the central regions of the galaxy).

The results from the tests are shown in Fig(s). 1 and 2 for the good resolution case, and Fig(s). 3 and 4
for poor resolution. From Fig. 1 we can observe that the reliability of the code is excellent in recovering the
surface brightness, scale length and position angle both of the bulges,
indicated by the open squares, and disks, indicated by the open circles. While there are some minor trends,
the agreement is always within the estimated fitting error delivered by the program.
The ellipticities of bulges and
disks are more difficult to estimate and are subject to larger errors. The
reason is that for a fixed value of ellipticity, within the formal error, we
can always find a set of the other photometric parameters that will result in a
final error close to the best model. Therefore, the final error is relatively
insensitive to an error in the intrinsic ellipticity of the two components.
Moreover, this effect tend to be worse when the bulge/disk ratio is larger than 1,
since in that case a change in the disk ellipticity practically has no impact
in the total error budget of the model. Therefore, great care should be
exercised when interpreting the ellipticity results. The ellipticity of the
most eccentric bulges may be underestimated, but again is within the errors. This ellipticity issue
may be easily corrected with the actions discussed in the above paragraph, by using, for instance, {\sc iraf}
tasks to have an initial value based on the observed image. The solid symbols in
Fig(s). 1d and 3d show how the code works in the cases when the ellipticities of bulge and disk
are treated as in real galaxies. The Pearson correlation coefficient R in these figures are relative only to
the solid symbols. In Fig. 2 we present
the results regarding the S\'ersic index and the bulge/disk ratios. For large
values of $n$ the adopted seeing can cause an important variation in the
estimated value. However, for the vast majority of galaxies, including Es, this
effect is not important since that index is bound to have a lower value close
to the de Vaucouleurs ($n=4$) model. The good agreement also found in Fig(s). 3 and 4,
where we have tested the impact of having low spatial sampling, is encouraging.

To verify the error produced in the residual images by the differences in the input and retrieved synthetic
models, we have subtracted the latter from the former in Fig. 5. The left panels show a case of good spatial
resolution while the
right ones show a case of poor resolution. They show that the discrepancy is restricted to
the central region, which has a size of the order of the seeing.
The right panels represent an especially bad fit
when the ellipticities were not treated as must be done in real galaxies. Such a problem may be easily
identified by the bipolar pattern left. The entire original synthetic galaxy occupies the
whole frame in the upper panels. The brightness and contrast levels displayed were chosen as to enhance the
differences.

Another interesting point regarding these tests is that the code is able to identify disks
which are up to 6 times fainter than the bulge (see Fig(s). 2 and 4), although this value depends somewhat on
the geometric differences between disk and bulge (in fact, our results below show that the code can be even
more sensitive). In any case, this result reinforces the {\sc budda} finding
of \citet{gad03} on the absence of disks in the SB0 galaxies NGC 4608 and 5701, since the bulge/disk ratio in
normal S0s is around 2 \citep{bin98}.

As noted above, the errors determined by the algorithm may be sometimes overestimated. This seems to be
especially true for the brightness and for the position angle. The $1\sigma$ error bars for the surface
brightness are typically around 0.75 mag. However, from Fig(s). 1 and 3, one sees that fiducial errors
may be considered to be around 0.25 mag. In these figures, there are no error bars associated with the
position angles because most of the time the algorithm is not able to estimate them reliably, a fact that
is related to the ellipticity issue discussed above. There are no error bars also related to the bulge/disk ratios
(Fig(s). 2 and 4), but this is because they are calculated via the model images.

\section{A Structural Analysis Study of 51 Galaxies}
In order to test {\sc budda} in real galaxies we have observed 51 galaxies at the Pico dos Dias Observatory
(PDO/LNA -- CNPq, Brazil). The only criterion these galaxies obey is to be brighter than around 14 in B, and
emphasis was put on Es and S0s. The CCD observations were done with a 24 inch telescope having a
focal ratio f/13.5, and using a thin back--illuminated CCD SITe SI003AB, with 1024
$\times$ 1024 pixels. The plate scale is 0.57 arcsec/pixel, resulting in a field
of view of approximately 10 $\times$ 10 arcmin. The CCD gain
was set at 5 e$^-$/ADU and the read--out noise on 5.5 e$^-$.
All objects were observed in the R passband of the Cousins system.
For each object, we have done 3 exposures of 300 seconds. The multiple
exposures aim to ease cosmic ray removal. The data were calibrated with a set of standard stars
from \citep{gra82} and corrected for atmosphere and Galactic extinction. The later correction
was done using the maps of \citet{sch98}. The standard processing of the
CCD data includes bias subtraction, flatfielding
and cosmetics. The first step in the sky subtraction was done by editing the combined
images, removing the galaxy and stars. After that step we
determined the mean sky background and its standard deviation ($\sigma$). Then,
we removed all pixels whose values were discrepant by more than 3 $\sigma$ from
the mean background. A sky model was obtained by fitting a linear surface to the
image, and this model was subtracted from the combined image. We finally removed
objects such as stars and H{\sc ii} regions. All these procedures were done
using the {\sc iraf} package. The observations were done in 6 nights in October 1997. Typical values
for the seeing and zero point error are, respectively, 0.8 to 1.2 arcsec and 0.02 to 0.05 mag.

Besides testing {\sc budda}, these data also allow us to perform a detailed structural analysis of galaxies.
Since our primary concern is to verify sub-structures in early--type galaxies, our sample contains mostly
Es and S0s, becoming also suitable to Fundamental Plane studies. However, our sample spans the whole
Hubble sequence, with some examples of Ss, which are useful to test {\sc budda} in galaxies with spirals, bars,
dust lanes and morphological peculiarities. Figure 6 shows the results for all galaxies.
For each of them we have obtained with {\sc budda} the structural
parameters and artificial images of the model galaxy, as well as of the model bulge and disk, separately. Using
the {\sc iraf} {\sc ellipse} task, we built the surface brightness profiles of the galaxy, of the model galaxy, and of
the bulge and the disk, as well as ellipticity, position angle and the b4 Fourier coefficient profiles for the galaxy and
for the model galaxy. The behavior of the b4 coefficient indicates how much the isophotes deviate
from a perfect ellipse. A positive value for b4 indicates isophotes with a disky component, whereas
a negative value indicates boxy isophotes. If b4 = 0, then the isophote is a perfect ellipse. This whole
procedure allows us to evaluate the quality of the fit and other structural parameters of the
galaxy. Moreover, we obtained residual images, subtracting from the galaxy the model images, to verify
the presence of sub-structures. In Fig. 6 we show, from left to right, and from top to bottom:

   \begin{itemize}
   \item Original Images, passed through a median filter and converted to surface brightness, with the
   isophotal map (with a 0.5 mag interval between each contour level)
   \item Total Residual Images, i.e., Original Image divided by Complete Model\footnote{Note that the
   total residual image is a ratio image, in which discrepancies are emphasized.}
   \item Disk Residual Images, i.e., Bulge Model subtracted from Original Image
   \item Bulge Residual Images, i.e., Disk Model subtracted from Original Image

   \item Surface Brightness profiles in the R--broadband, in magnitudes per squared arcsecond, as a function
   of the radius, in pixels, where

   \begin{itemize}
   \item points with error bars represent the Original Image
   \item the short--dashed line represents the Disk Model
   \item the long--dashed line represents the Bulge Model
   \item the full line represents the Complete Model
   \item points with full line represent the Total Residual profile (original image $-$ complete model
   $+$ 28), in magnitudes per squared arcsecond
   \end{itemize}

   \item ellipticity, position angle, and b4 Fourier coefficient profiles, where

   \begin{itemize}
   \item points with error bars give values from Original Images
   \item the full line represents the Complete Model
   \end{itemize}

   \end{itemize}

Through a careful analysis of Fig. 6, one can identify and study several features which would otherwise
remain hidden in the direct galaxy image. Some points must be stressed concerning the analysis of residual
images. In the computer, one may always change the brightness and contrast levels to carefully identify
what is real and what is spurious. To make the figures, these levels are chosen to highlight features one
wants to point out. Due to seeing effects the first few pixels in the nucleus are always brighter than
the model and this excess extends to a few more pixels because there are too few data points for the
model to be accurate. Thus, analysis in the {\em very} inner parts of the galaxy is always dubious.
A level of subjectivity still remains in this kind of analysis, and because of that, the existence of any
sub-structure may be confirmed also through the analysis of the radial profiles.
For instance, an inner disk or bar may reveal itself in surface brightness radial profiles as a bump.
Some other signatures are described in the following subsections.

Regarding our results, a first point that should be mentioned is that the broadest classification
one may do with our sample is to distinguish between galaxies with only one spheroidal component, and galaxies
that contain also an exponential disk. Behind this apparently banal statement, there are elliptical galaxies,
which {\em should} be pure spheroidals, but contain inner disks, and lenticular galaxies, which {\em should}
have a disk, but do not show any discernible disk, either because of wrong classification or secular evolution
(in the case of 1 SB0, namely NGC 2217). Moreover, we find several other cases of wrong morphological
classification and many galaxies with hidden sub-structures. We discuss these generic cases in the following
subsections. Table 1 presents our results, including the structural parameters found for each galaxy.

We have made an extensive search in the literature for similar estimates for the galaxies in our sample,
but it turned out that only to one galaxy a 2D decomposition code has been applied before in the R
broadband. This is NGC 7177, studied also by \citet{dej96a}. Unfortunately, however, our bulge parameters
can not be compared since de Jong a priori assumed an exponential bulge for this galaxy, while our
results show that its bulge is better described by a S\'ersic profile with $n=3.52$. On the other hand,
the disk parameters may be compared. The agreement shown is good indeed: while de Jong's results
point to $I_{0d}=18.73$ R mag arcsec$^{-2}$ and $h=15.99$ arcsec, {\sc budda} estimates
$I_{0d}=18.58$ R mag arcsec$^{-2}$ and $h=12.97$ arcsec. It is interesting that, at least in this case,
using a different bulge luminosity profile did not alter significantly the estimate of the disk parameters.
Moreover, the fact that de Jong also used a model for the bar of this galaxy indicates that {\sc budda}
was able to distinguish the contribution to the total luminosity from the bulge, disk and bar, without
assuming any luminosity distribution for the bar, which is very encouraging.

\subsection{Inner Disks in Ellipticals}

There are 23 galaxies in our sample classified as Es in the RC3 \citep{dev91}. From these, only 9 seem to us
to be pure spheroids, since the fit is excellent with only the spheroidal component and the residual images
do not show any sub-structure (except in the nucleus, where seeing prevents any analysis). Prototypical
examples are NGC 822, 6578, 6958, 7507 and 7619. Others that fall in this category are: IC 4842,
4943, NGC 720, and 1199. But even on some of these examples some doubt may be cast
regarding the presence of any more complex morphology. \citet{bur01} suggest that {\em all} galaxies have
disks but 20\% of them are not forming stars and a small fraction are too faint.

Eight galaxies classified as Es do show inner discs. These may be identified directly by the code, which
in this case finds the exponential profile. This is the case for e.g. NGC 1052 and 7192. In some cases,
however, the disk is too faint or too small to be recognized by {\sc budda} but it appears clearly in the
residual images. One must bear in mind that these inner disks are not necessarily exponential disks, which is
what our code tries to fit. Examples of this case are ESO 115-G008, PGC 64863 and NGC 821. Other signatures
of the inner disk are disky (b4 $>$ 0) inner isophotes and a distinct behavior of the ellipticity and position angle
radial profiles from the inner to the outer regions of the galaxy. Disky isophotes are a signature of inner disks
because it was found that Es with disky isophotes rotate faster \citep{kor89}. However, the isophote being disky
only means that there is an excess of light along the major and minor axes of the galaxy, which makes it look pointy.
Bars seen close to or face--on also may produce disky isophotes. When b4 $<$ 0, the isophotes are boxy.
This means that the galaxy has a deficiency of light along its main axes, which causes a rectangular shape.
Boxy isophotes are also found in some Es, which have negligible rotation \citep{kor89}. Finally, boxy isophotes
are also related to bars, but when seen edge--on \citep{bin98}. Other Es with inner disks are:
NGC 1209, 1700 and 7196.

The remaining 6 galaxies are examples of bad morphological classification and will be discussed below.
In all these galaxies we have found a large exponential disk which puts them in the lenticular class. In the
misclassification section there are also galaxies classified as S0s, but that are actually Es, some of them
with inner disks or complex morphology.

\subsection{Sub-structure in Lenticular Galaxies and another SB0 without Disk}

There are 20 galaxies in our sample classified as S0s in the RC3. Eleven are genuine lenticulars, in which
{\sc budda} determined the disk parameters, except in one, namely NGC 2217. This galaxy apparently
has a very strong bar that could have destroyed the pre-existent disk within the bar length, and may
be another case as those (NGC 4608 and NGC 5701) discovered by \citet{gad03}; see also \citet{ath03}.
Note, however, that, while NGC 4608 and NGC 5701 have indeed strong bars, as denote by their Fourier $m=2$
component \citep{gad03}, it remains to be seen that the bar in NGC 2217 is really strong. As pointed out
by \citet{but01}, bar strength can be defined in terms of tangential forces relative to mean radial forces, and
the latter are certainly significant in galaxies with large bulges, such as NGC 2217. Note, furthermore, that the
existence of conspicuous rings outside the bar in these galaxies may be an indication of the existence of an
outer disk, as the bar is only able to consume the disk within it.

Three of the eleven genuine lenticular galaxies are strongly
barred and there are also three more strongly barred S0/a galaxies in our sample.
In many of these systems one encounters interesting sub-structures which do not appear in their direct images.
In some cases, structural peculiarities may be seen and recognized in a careful analysis of
the direct images, but are certainly better defined in the residual ones.
In NGC 1326 and 2217, the inner ring surrounding the bar, already identified by \citet{dev64},
appears more clearly in the residual image \citep[for a detailed analysis of NGC 1326, see][]{but98}.
Surprising dust
lanes are visible in NGC 1316 and 1947, while the residual image of NGC 7280 displays a strong lens. An
extremely interesting case is NGC 128. This edge--on galaxy, which is a prototype of peanut-shaped bulges was
fitted by our code using its edge--on capabilities (see Sect. 2). The other objects which fall in this category
are: IC 5250A, NGC 467, 2293, 6673, and 7377. The bulge/disk luminosity ratios for the lenticulars in our sample
are around 2 with some noticeable deviation cases, in agreement with \citet{bin98}.

In the remaining 9 galaxies we haven't found a large disk. These are clearly Es rather than S0s and will be discussed
as misclassification cases below.

\subsection{Wrong Morphological Classification}

To distinguish between an elliptical and a lenticular galaxy is a hard task with visual inspection only, especially
for faint and/or distant galaxies. Thus confusion between Es and S0s is common and may badly interfere
in studies of e.g. scaling laws and the origin of the Hubble sequence. This is a point where codes like {\sc budda}
may play an important role given their higher reliability. In our sample, 6 Es were found to have large disks which
means they are actually S0s: IC 1459, NGC 1172, 1549, 2305, 7145, and 7778. In 3 of them (IC 1459, NGC
1549 and 7145) a lens is clearly visible in the residual images.

On the other hand, 9 galaxies classified as S0s were found to have no large disk, which puts them in the E bin.
As was also the case for some genuine ellipticals, some of these 9 systems show a morphology more
complex than a pure spheroid. Prominent cases are the disturbed morphology of NGC 7252, the {\em inner} disk
in NGC 2271, and the lens in NGC 1553, whose ratio image displays also an inner spiral in the center.
The last case is especially intriguing since lenses are common
in S0s. Considering the results from our code, either this is a rare case of a lens in an elliptical galaxy, or
(more likely) {\sc budda} was not able to find the disk in this lenticular galaxy simply because its disk has not the
common properties of exponential disks generally found in other disk galaxies. Its disk could possibly be
inner truncated or have a flat inner profile (type II profile). The fact that its lens is also very bright and has
an unusual appearance may
have prevented our code to naturally find reliable estimates for the disk structural parameters.
Indeed, {\sc budda} finds a disk in this galaxy, but that is a {\em very} faint disk with unusually
high error estimates. It is worth noting that, as the position angle and the ellipticity of the lens are
very similar to those of the bulge (and possibly the disk), it is very hard to disentangle the
contribution of the lens to the total luminosity of this galaxy without a priori assumptions.
NGC 1553 is certainly a case in which a more detailed analysis is in order.
The other misclassified Es are: IC 1919, NGC 541, 2205, 6849, 7289, and PGC 67633.

It is interesting to point out that from the 43 Es and S0s galaxies in our sample (as considered in the RC3)
15 were shown to be in the wrong morphological bin, i.e., $\approx$ 35\%. Moreover, it seems (not surprisingly)
that it is equally easy to misclassify an elliptical as an S0 and the other way around.

\subsection{Sub-structure in Spiral Galaxies}

Eight galaxies in our sample have a morphological class later than S0 according to the RC3. In many of them
hidden sub-structures emerge from the residual images. We point out: the bar and the strong spiral arms in
NGC 1079; the amazing case of NGC 1291 with the nuclear bar and inner ring\footnote{This ring may be
actually a lens. Since a lens is like a ring of low surface brightness amplitude
\citep[see][]{but96}, when a declining bulge
and disk background is removed the lens then becomes very ringlike. In this case this may be
particularly true given the broad appearance of this structure (but note that it is connected with the
ansae at the ends of the bar).} that only show out clearly in the
residual images [but note that \citet{dev75} already recognized these features];
the bar and inner ring of NGC 7177; the bar and the unusual spiral arms of NGC 7371, and
finally the weird spiral structure of NGC 7824. This last case is especially puzzling because, in spite of
the spiral structure, a model containing only a bulge is able to fit the surface brightness profile very well.
Again, this may be a case where the disk can not be modelled as it is traditionally done, i.e., with a exponential
disk. The inner spirals seem to be detached from the outer ones. And the latter have an unusual broad
appearance. Bars also reveal themselves in radial profiles of ellipticity (monotonically growing) and position
angle (constant). The other spirals in our sample are: NGC 1637, 7070A, and 7171.

It is worth noting that in cases like NGC 1079 and NGC 1291, where our images do not sample the outermost
parts of the galaxy, a better result for the disk model might be achieved with deeper and larger images
(especially because, in these examples, the fits are affected by the outer ring feature). Also, in NGC 7171,
the bulge looks overly important at large radii, which may be causing the bright halo around the ratio image.
This is a case in which a more realistic model may be accomplished at the expenses of some a priori
assumptions, like e.g., keeping the S\'ersic index constant or assuming an outer truncation for the bulge.

\subsection{Nuclear Activity and Companions}

There are 7 galaxies in our sample which are classified as AGNs at some level (e.g., Liner, Seyfert) in the
Nasa Extragalactic Database\footnote{See http://nedwww.ipac.caltech.edu/}. They are: IC 1459, NGC 1052,
1316, 1326, 1553, 7177, and 7280. There are also 10 galaxies which are classified as multiple systems
in the Leda Extragalactic Database\footnote{See http://leda.univ-lyon1.fr/} or show a clear companion in our
images. This does not mean galaxies in groups or clusters and does not exclude systems that are only
due to projection. It only means that these galaxies have a close companion in the plane of the sky.
These are: ESO 115-G008, IC 5250A, NGC 128, 467, 1052, 2293, 7070A, 7196, 7252, and PGC 67633.
The only galaxy which is a member of both sub-samples is NGC 1052. We have not found any clear tendency
in these 2 groups of galaxies that makes them distinct from the whole sample. Only two wrong classification
cases have companions, which means that these cases may not be caused by some unknown tidal
process that could slightly change the morphology of a galaxy.

\section{Concluding Remarks}

We have thoroughly presented {\sc budda}, a code devoted for bulge/disk decomposition in galaxy images. As was
shown in details above, its development and performance tests show that it is a very suitable tool for the structural
analysis of galaxies. The parameters for the bulge and disk models can be retrieved with great confidence
as well as the residual images obtained from them. Its simplicity to use and its broad range of applications add
to its qualities.

Applying {\sc budda} on our sample of 51 galaxies we could check how it works concerning its many potential
applications. One obvious use of the code is to obtain bulge and disk structural parameters to study e.g. the
Fundamental Plane and other scaling relations. Such a study regarding the galaxies in our sample will
appear in a forthcoming paper. Such relations may also be strengthened if wrong morphological classification
cases are identified, which is reliably done with our code, as we showed in Sect. 4.3. Galaxies classified as
Es, but actually being S0s, which was the case for 6 of our galaxies, may add to the spread in scaling
relations. Obviously, the opposite is also true. In the FP, for instance, S0s have generally lower mass-to-light
ratios than Es \citep{ben92}.

Another use of {\sc budda}, in which we've found it to be very powerful, is to reveal hidden sub-structures,
especially in the inner regions of galaxies. In Sect(s). 4.1 and 4.3 we revealed inner disks in many elliptical galaxies,
which may be the result of a recent merger. Or are they related to galaxy formation processes? {\sc budda} has
found also several examples of inner spirals, inner and nuclear rings, as well as inner and secondary bars in S0s
and Ss. In some cases a hint of them shows up in the direct images or in the profiles, but in others there is no
a priori sign of them at all. These sub-structures, which appear conspicuously in the residual images, are relevant
for e.g. dynamical studies, since they are related to non-circular orbits and Lindblad resonances. Moreover,
these sub-structures may possibly be relevant to the long standing problem of fueling active galactic nuclei
\citep{com01}, and may be directly studied in the residual images.

\acknowledgments
We thank the Pico dos Dias Observatory staff for support during the observing runs. Financial support from FAPESP
grants 99/07492-7 and 00/06695-0 is acknowledged. We would like to thank the anonymous referee for many
enlightening comments and helpful suggestions.

\newpage

\figcaption[f1.eps]{Comparison between the parameters of the 41 synthetic galaxies as retrieved
by {\sc budda} in the case of good spatial sampling: (a) central surface brightness of the disk and
effective surface brightness of the bulge in arbitrary magnitudes; (b) characteristic radius of the
disk and effective radius of the bulge in arbitrary units; (c) position angle, and (d) ellipticity.
Although in (a) and (b) the units are arbitrarily chosen, all values displayed here are similar to those found
for nearby observed galaxies.
Circles are for the disc and squares are for the bulge parameters. The error bars were not considered
for clarity reasons but a typical error bar is shown in each panel. Note that the typical error bar
considered is as determined by {\sc budda}, which is not necessarily similar to the scatter in this figure.
For instance, in (a) it is possible to see that our code typically overestimates the errors for the
intensities. In (c) there are no error bars, for the code typically does not find a meaningful error estimate
for the position angles, since variations in these parameters do not cause significant changes in $\chi^2$.
The Pearson correlation coefficient R of each test is also indicated. For (d), R is relative only to
the solid symbols. These refer to those tests in which the first guesses for the ellipticities were
not random, but determined from the results of the {\sc ellipse} task in {\sc iraf}, as is done
for real galaxies. \label{fig1}}

\figcaption[f2.eps]{Same as Fig. 1 but for (a) S\'ersic index, and (b) the bulge/disk ratios in luminosity
(circles) and size (squares). \label{fig2}}

\figcaption[f3.eps]{Same as Fig. 1 but for poor spatial sampling. Typical error bars are as in the good
sampling case (see Fig. 1). \label{fig3}}

\figcaption[f4.eps]{Same as Fig. 2 but for poor spatial sampling. Typical error bars are as in the good
sampling case (see Fig. 2). \label{fig4}}

\figcaption[f5*.eps]{Examples of the discrepancy between a synthetic galaxy built with {\sc bmodel} and the one
retrieved by {\sc gmodel} within {\sc budda}. The upper left panel shows a case of good spatial resolution while the
upper right one shows a case of poor resolution. The last is an example of a specially bad fit. Even so, the
differences are restricted to the central region, since the entire original galaxy occupies the
whole frame. The brightness and contrast levels displayed were chosen as to enhance the
differences. The lower panels show in details the central region of each of the former images,
respectively. \label{fig5}}

\figcaption[f6*.eps]{Results of the {\sc budda} algorithm applied in R-band CCD
images for all the 51 galaxies in our sample. At the top it is shown the name of the galaxy, its
morphological type according to the RC3 and the size of the images in arcseconds. See
the text for further explanations. The remaining figures are available in the electronic edition of the Journal.
The printed edition contains only a sample. \label{fig6}}

\epsscale{0.85}
\newpage
\plotone{f1.eps}

\newpage
\plotone{f2.eps}

\newpage
\plotone{f3.eps}

\newpage
\plotone{f4.eps}

\newpage
\plotone{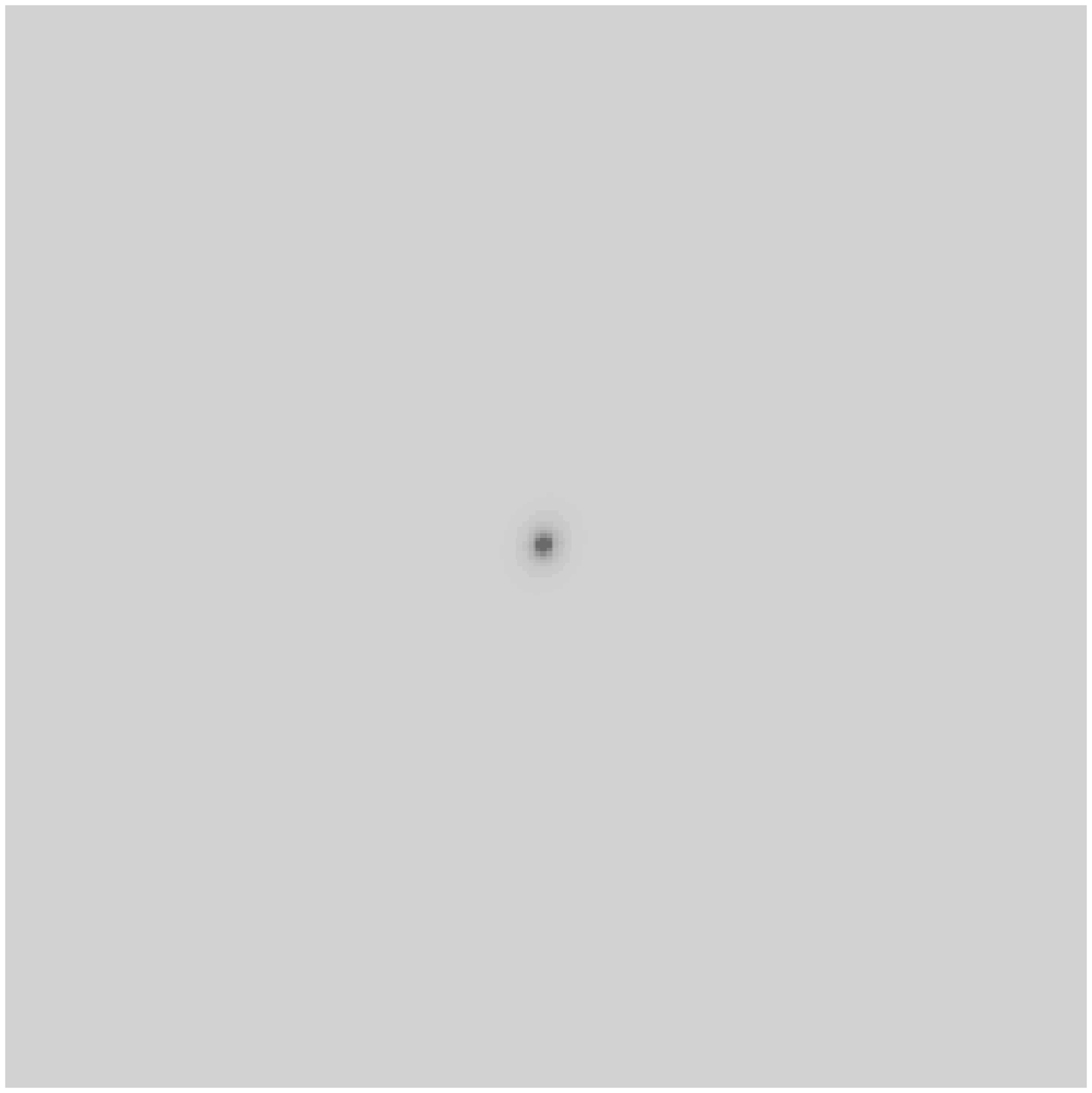}
\newpage
\plotone{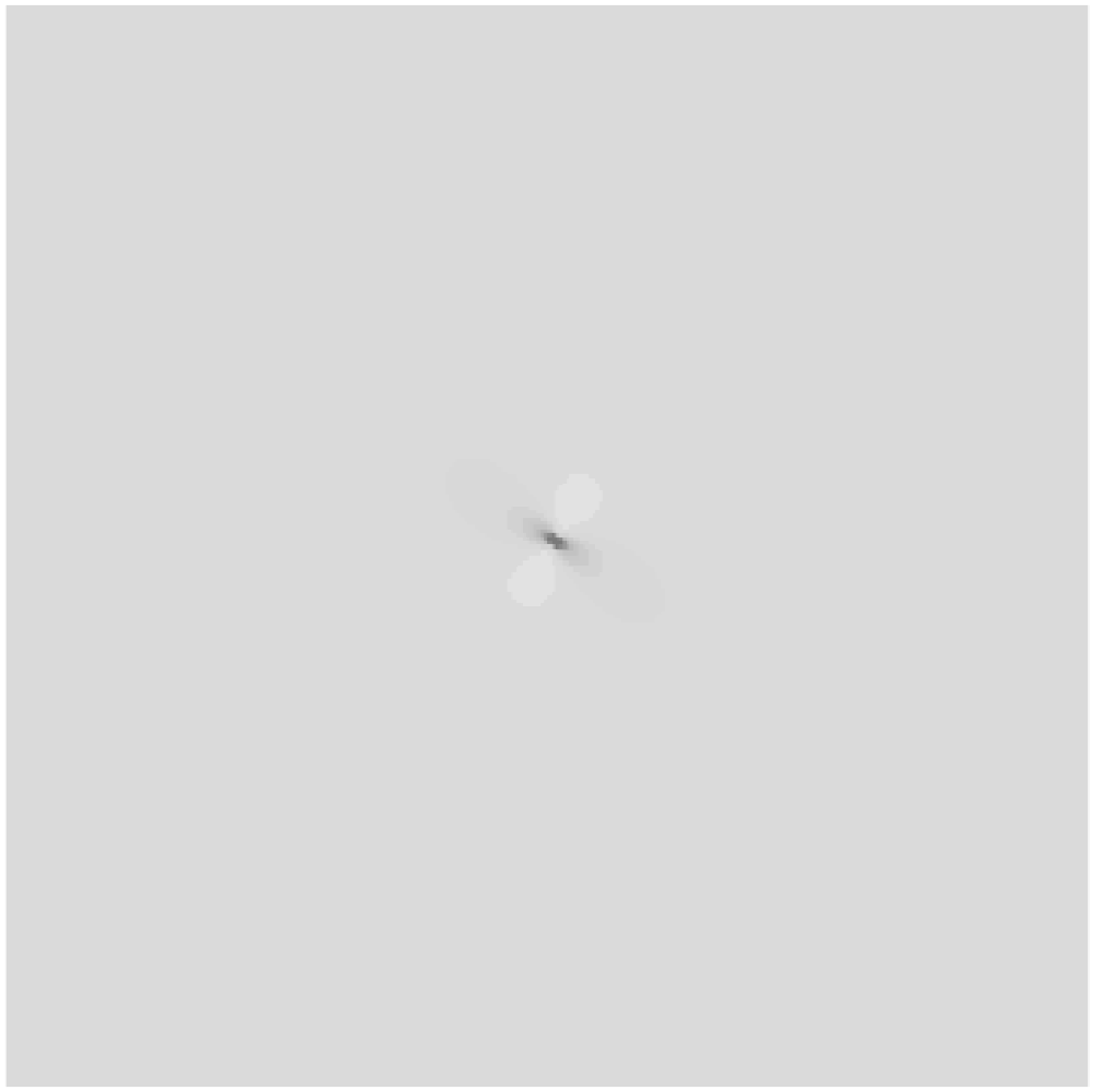}
\newpage
\plotone{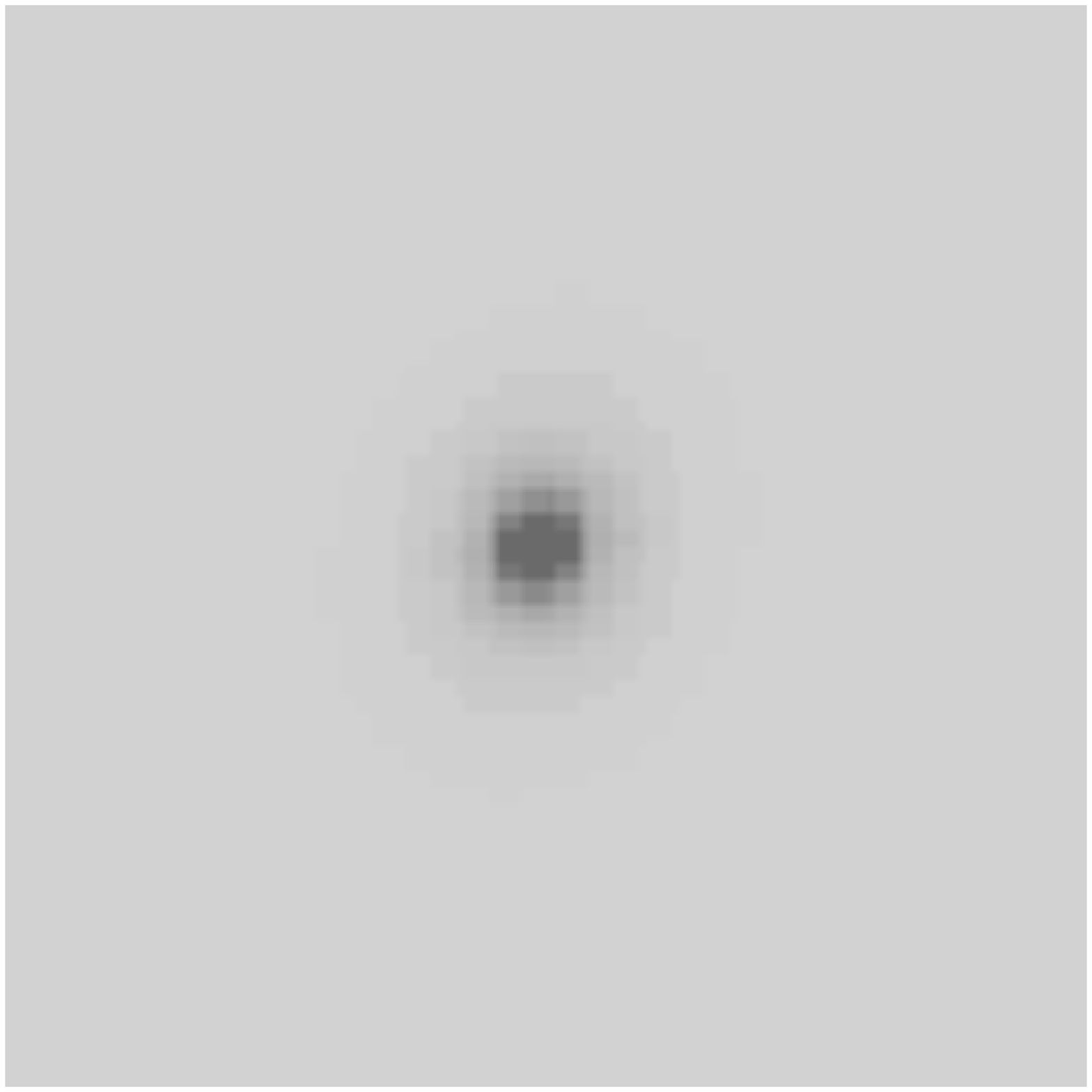}
\newpage
\plotone{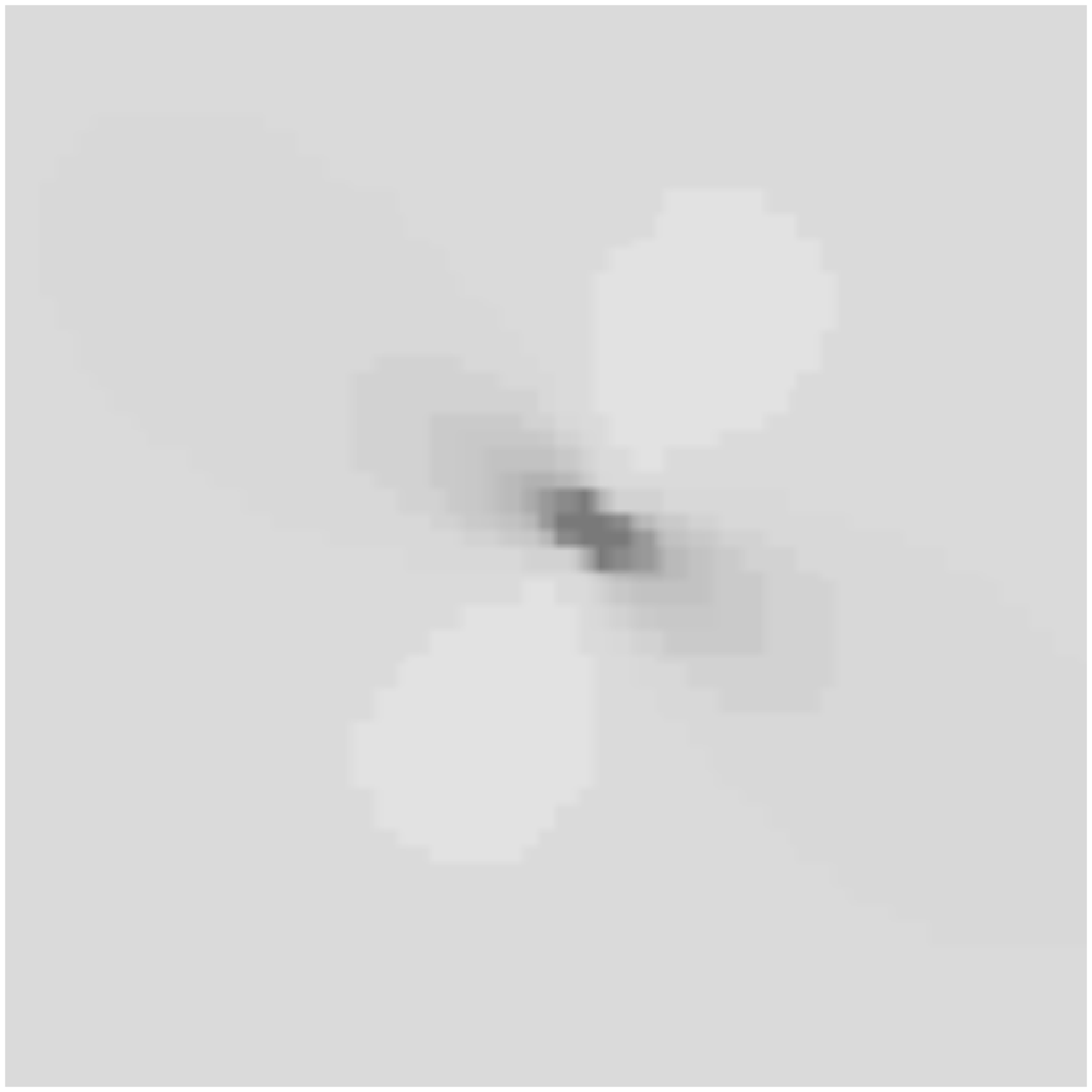}

\newpage
\plotone{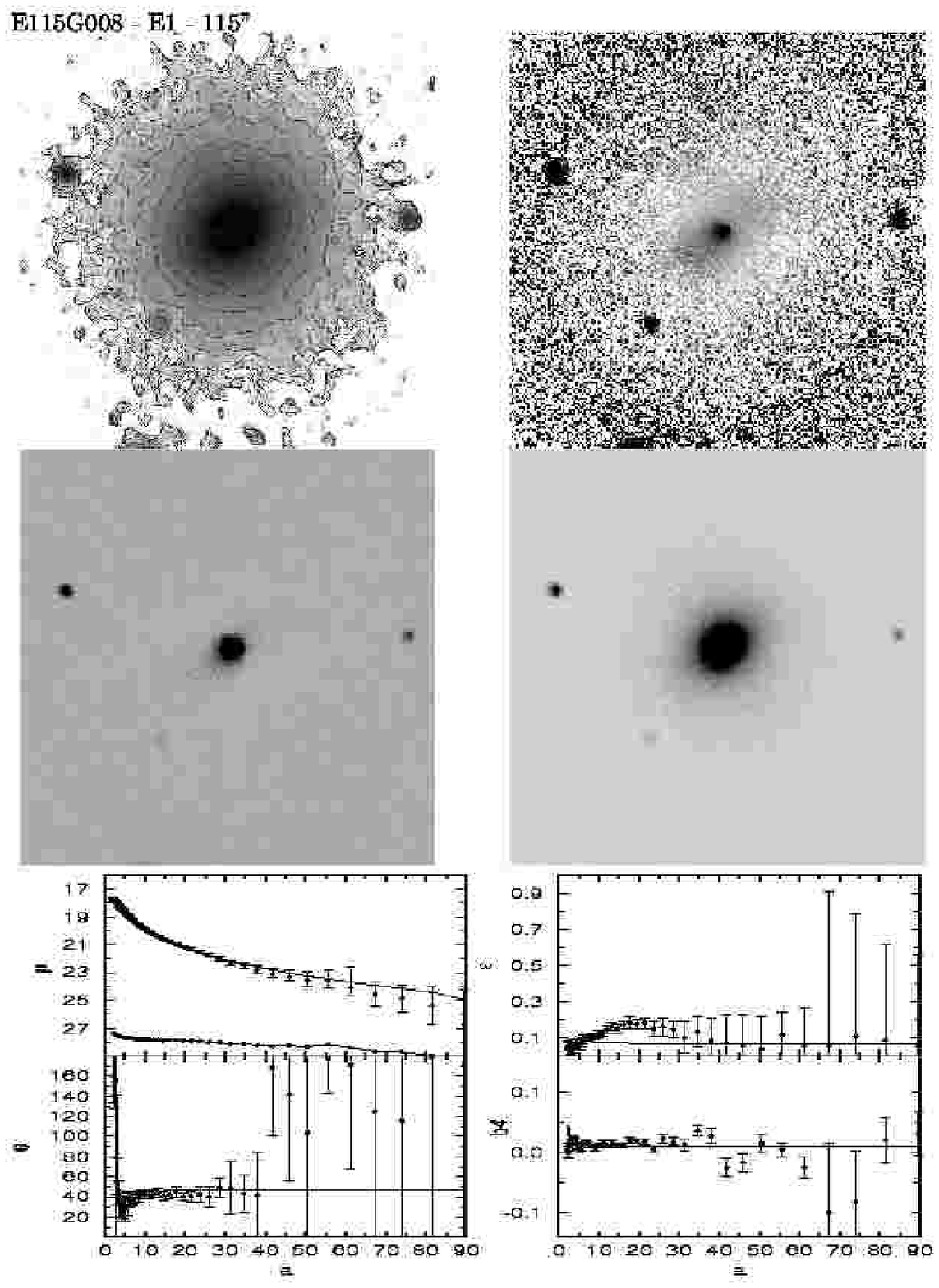}
\newpage
\plotone{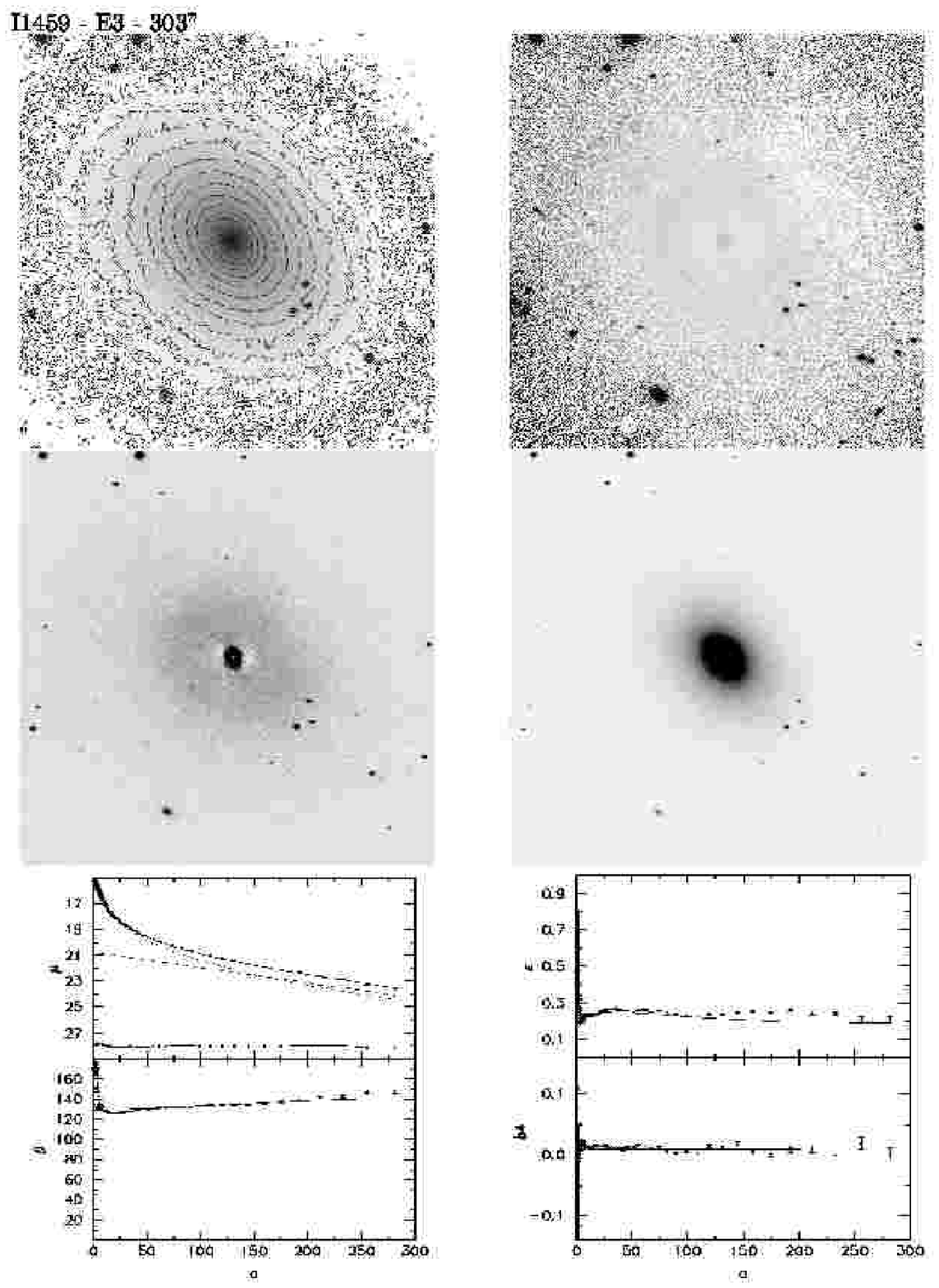}
\newpage
\plotone{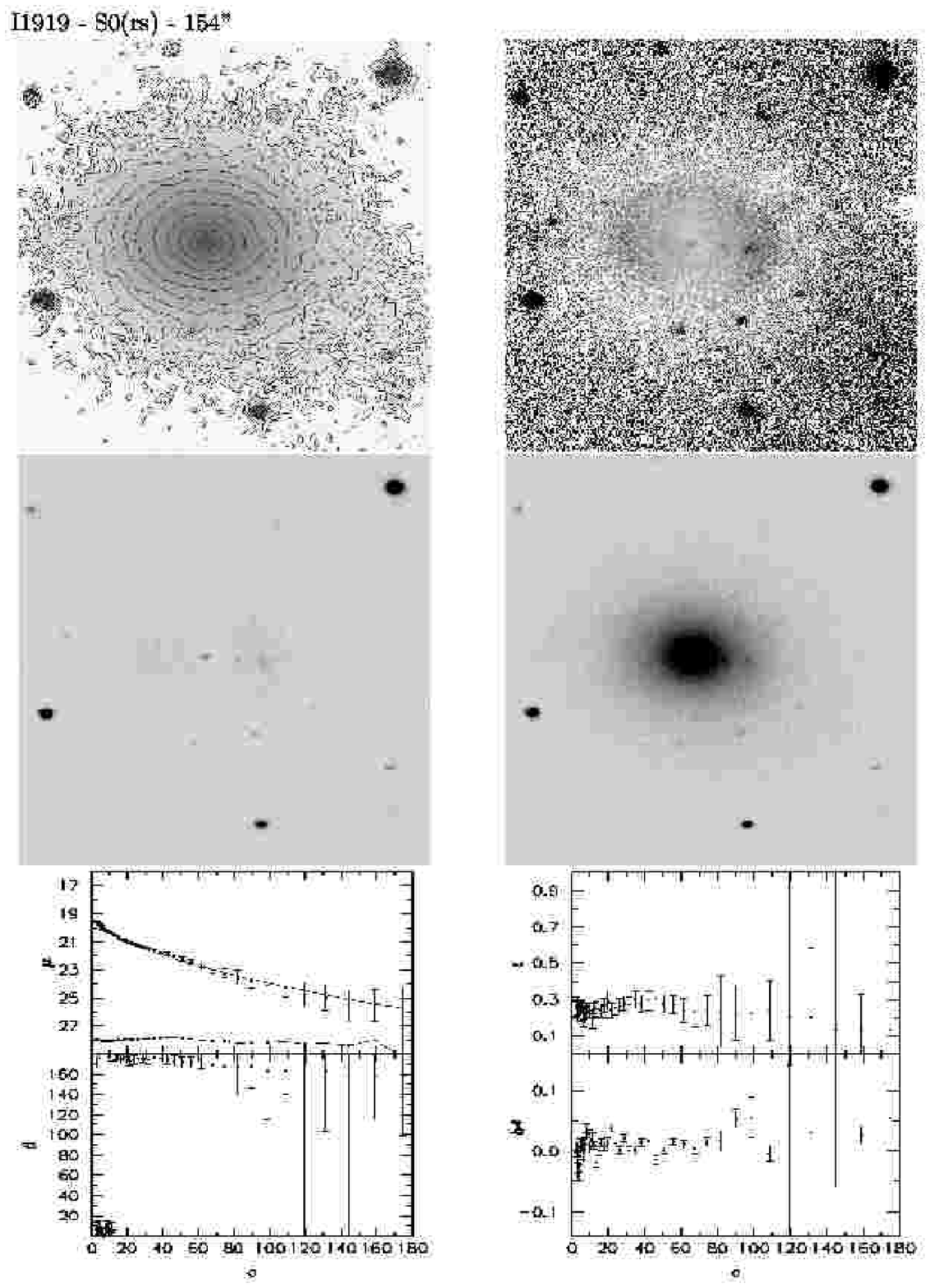}
\newpage
\plotone{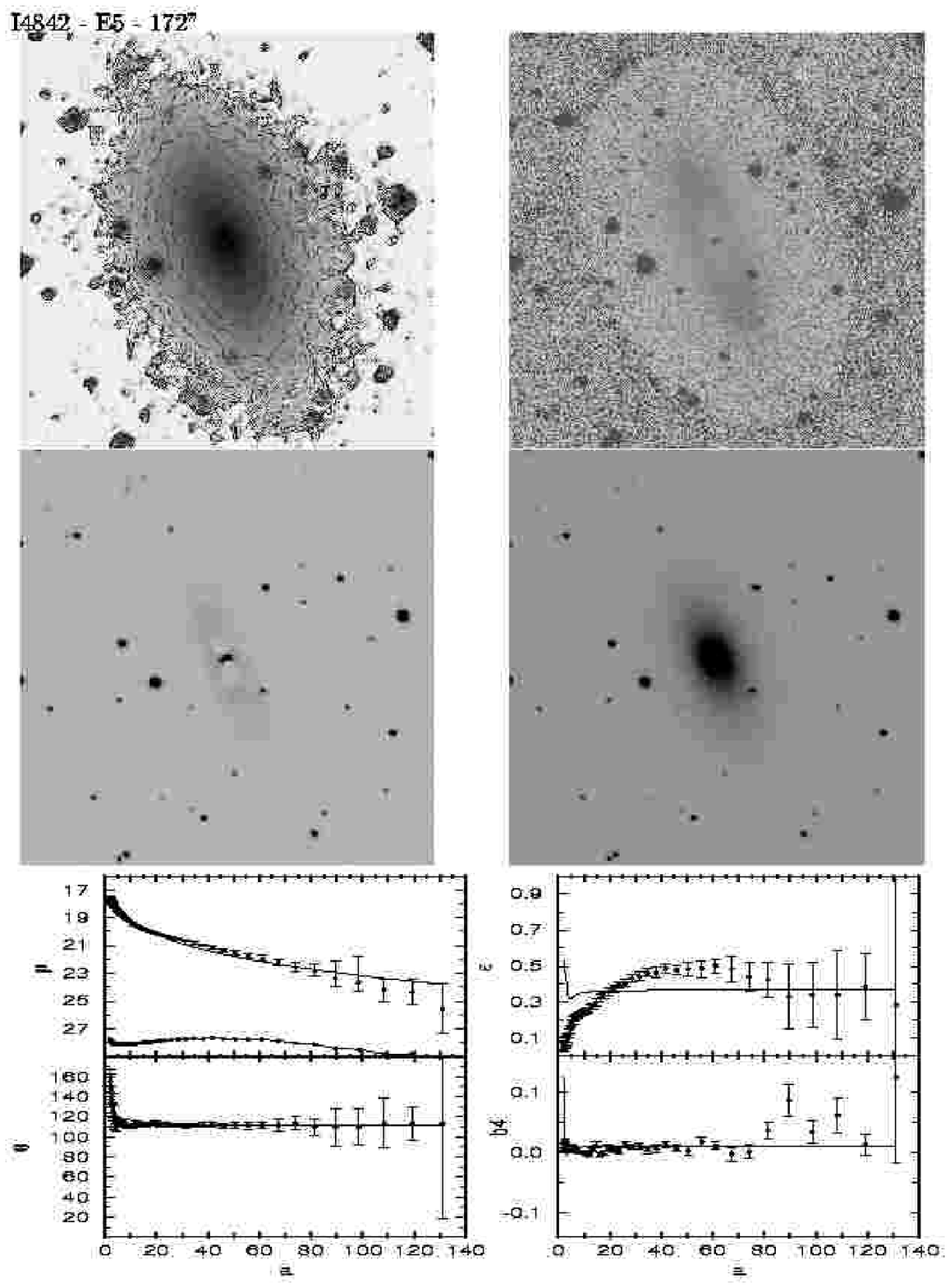}
\newpage
\plotone{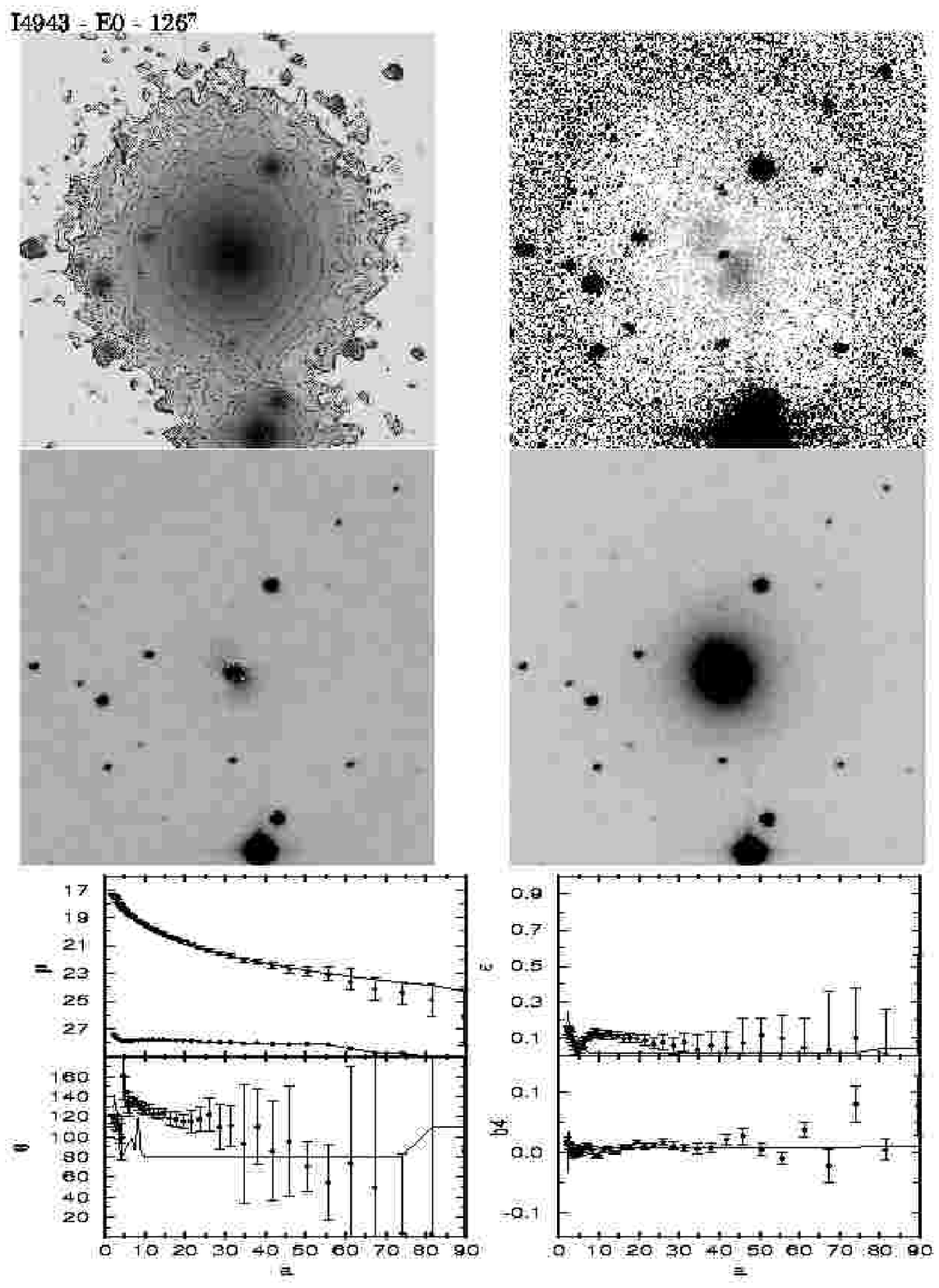}

\clearpage
\begin{deluxetable}{cccccccccc}
\rotate
\tabletypesize{\scriptsize}
\tablecaption{Results from the structural analysis for 51 galaxies. \label{tbl-1}}
\tablewidth{0pt}
\tablehead{\colhead{Galaxy} & \colhead{T} & \colhead{B/D} & \colhead{B/D}
& \colhead{$I_{0d}$} & \colhead{$h$\tablenotemark{a}}
& \colhead{$I_{eb}$} & \colhead{$r_{eb}$\tablenotemark{a}} & \colhead{$n$\tablenotemark{a}}
& \colhead{Comments} \\
\colhead{} & \colhead{(RC3)} & \colhead{(Luminosity)} & \colhead{(Size)} &
\colhead{(R mag arcsec$^{-2}$)} & \colhead{(arcsec)} &
\colhead{(R mag arcsec$^{-2}$)} & \colhead{(arcsec)} &
\colhead{} & \colhead{}}
\startdata
ESO115G008 & E1 & \dots & \dots & \dots & \dots & 21.02 & 18.30$\pm$1.10 & 4.67$\pm$0.69 & inner disk \\
IC1459 & E3 & 2.4 & 1.9 & 19.54 & 52.66$\pm$8.46 & 18.56 & 28.23$\pm$1.70 & 3.72$\pm$0.59 & trueS0+lens \\
IC1919 & S0(rs) & \dots & \dots & \dots & \dots & 21.31 & 31.12$\pm$1.56 & 1.87$\pm$0.24 & trueE \\
IC4842 & E5 & \dots & \dots & \dots & \dots & 21.65 & 49.04$\pm$2.71 & 4.79$\pm$0.57 & \dots \\
IC4943 & E0 & \dots & \dots & \dots & \dots & 20.96 & 21.96$\pm$0.49 & 5.28$\pm$0.32 & \dots \\
IC5250A & S0 & 0.43 & 1.10 & 20.15 & 67.26$\pm$22.85 & 20.59 & 15.09$\pm$2.60 & 4.58$\pm$2.09 & \dots \\
NGC0128 & S0 & 3.51 & 2.88 & 19.04 & 20.22$\pm$3.60 & 18.94 & 19.94$\pm$1.58 & 2.20$\pm$0.46 & peanut \\
NGC0467 & S0 & 3.62 & 1.48 & 20.95 & 34.17$\pm$3.64 & 20.77 & 30.99$\pm$0.50 & 3.54$\pm$0.13 & \dots \\
NGC0541 & S0 & \dots & \dots & \dots & \dots & 21.33 & 35.84$\pm$1.54 & 3.93$\pm$0.38 & trueE \\
NGC0720 & E5 & \dots & \dots & \dots & \dots & 20.01 & 58.20$\pm$2.17 & 3.86$\pm$0.34 & \dots \\
NGC0821 & E6 & \dots & \dots & \dots & \dots & 20.78 & 56.56$\pm$3.77 & 4.42$\pm$0.67 & inner disk \\
NGC0822 & E3 & \dots & \dots & \dots & \dots & 20.66 & 18.69$\pm$0.59 & 4.34$\pm$0.36 & \dots \\
NGC1052 & E4 & 13.83 & 1.70 & 18.81 & 19.27$\pm$2.37 & 20.02 & 47.76$\pm$0.85 & 4.75$\pm$0.21 & inner disk \\
NGC1079 & SAB0/a(rs) & 100.24 & 1.89 & 21.53 & 13.40$\pm$22.54 & 20.78 & 50.86$\pm$2.26 & 3.82$\pm$0.41 & \dots \\
NGC1172 & E3 & 2.81 & 1.50 & 21.60 & 92.91$\pm$63.22 & 21.14 & 31.79$\pm$2.27 & 5.02$\pm$0.76 & trueS0 \\
NGC1199 & E3 & \dots & \dots & \dots & \dots & 20.41 & 35.54$\pm$0.92 & 3.89$\pm$0.24 & \dots \\
NGC1209 & E6 & \dots & \dots & \dots & \dots & 20.63 & 51.37$\pm$4.38 & 4.80$\pm$0.90 & inner disk \\
NGC1291 & SB0/a(s) & 2.63 & 1.48 & 17.75 & 35.35$\pm$3.14 & 19.56 & 55.88$\pm$2.58 & 4.63$\pm$0.51 & inner bar/ring \\
NGC1316 & SAB0(s) & 4.51 & 2.13 & 19.45 & 71.02$\pm$5.63 & 19.69 & 89.23$\pm$2.29 & 4.67$\pm$0.28 & dust lane \\
NGC1326 & SB0(r) & 4.11 & 1.73 & 19.91 & 32.63$\pm$7.59 & 20.02 & 38.54$\pm$2.23 & 4.89$\pm$0.71 & inner ring \\
NGC1549 & E0 & 9.52 & 1.45 & 19.47 & 27.90$\pm$6.10 & 19.70 & 52.25$\pm$2.22 & 4.42$\pm$0.44 & trueS0+lens \\
NGC1553 & S0 & \dots & \dots & \dots & \dots & 20.28 & 81.01$\pm$6.86 & 5.24$\pm$0.90 & trueE+lens \\
NGC1637 & SABc(rs) & 0.24 & 0.63 & 18.91 & 36.56$\pm$4.49 & 20.31 & 18.96$\pm$6.11 & 3.18$\pm$1.97 & \dots \\
NGC1700 & E4 & 22.46 & 1.68 & 18.75 & 9.61$\pm$1.04 & 20.05 & 36.50$\pm$0.34 & 3.89$\pm$0.08 & inner disk \\
NGC1947 & S0 & 9.44 & 1.44 & 22.62 & 33.03$\pm$12.94 & 22.17 & 49.58$\pm$2.08 & 2.81$\pm$0.23 & dust lane \\
NGC2205 & SAB0(rs) & \dots & \dots & \dots & \dots & 22.27 & 23.62$\pm$0.94 & 4.43$\pm$0.44 & trueE \\
NGC2217 & SB0(rs) & \dots & \dots & \dots & \dots & 22.24 & 59.01$\pm$7.15 & 4.40$\pm$1.48 & no disk \\
NGC2271 & SAB0 & \dots & \dots & \dots & \dots & 22.00 & 28.94$\pm$1.02 & 2.81$\pm$0.25 & trueE+inner disk \\
NGC2293 & SAB0(s) & 1.41 & 1.91 & 21.87 & 72.14$\pm$17.57 & 21.47 & 28.67$\pm$2.46 & 4.06$\pm$0.94 & \dots \\
NGC2305 & E2 & 2.80 & 1.84 & 22.49 & 44.10$\pm$14.73 & 21.03 & 19.75$\pm$1.09 & 2.99$\pm$0.45 & trueS0 \\
NGC6578 & E2 & \dots & \dots & \dots & \dots & 20.80 & 43.11$\pm$2.58 & 4.84$\pm$0.72 & \dots \\
NGC6673 & SAB0(s) & 7.92 & 2.06 & 20.69 & 33.72$\pm$34.14 & 20.19 & 34.60$\pm$3.16 & 5.07$\pm$1.11 & \dots \\
NGC6849 & SB0 & \dots & \dots & \dots & \dots & 21.73 & 51.57$\pm$2.62 & 4.12$\pm$0.42 & trueE \\
NGC6958 & E2 & \dots & \dots & \dots & \dots & 20.18 & 36.98$\pm$2.17 & 4.49$\pm$0.68 & \dots \\
NGC7070A & Irr & 1.26 & 1.64 & 19.75 & 20.43$\pm$2.73 & 20.92 & 22.65$\pm$2.04 & 2.74$\pm$0.48 & dust lane \\
NGC7145 & E0 & 4.65 & 1.44 & 20.88 & 41.37$\pm$7.14 & 21.06 & 44.35$\pm$1.25 & 4.88$\pm$0.29 & trueS0+lens \\
NGC7171 & SBb(rs) & 0.44 & 1.12 & 20.18 & 41.28$\pm$8.00 & 21.25 & 40.57$\pm$10.18 & 2.17$\pm$1.38 & \dots \\
NGC7177 & SABb(r) & 7.55 & 1.63 & 18.58 & 12.97$\pm$4.07 & 20.19 & 34.89$\pm$2.86 & 3.52$\pm$0.68 & inner ring \\
NGC7192 & E0 & 118.52 & 1.44 & 21.16 & 7.98$\pm$10.44 & 20.61 & 39.05$\pm$1.22 & 4.06$\pm$0.30 & inner disk \\
NGC7196 & E2 & \dots & \dots & \dots & \dots & 20.05 & 30.65$\pm$1.45 & 4.05$\pm$0.48 & inner disk \\
NGC7252 & S0 & \dots & \dots & \dots & \dots & 19.92 & 19.72$\pm$0.99 & 4.03$\pm$0.55 & trueE? \\
NGC7280 & SAB0(r) & 2.96 & 2.07 & 20.34 & 27.47$\pm$4.94 & 20.76 & 28.78$\pm$1.41 & 5.32$\pm$0.67 & lens \\
NGC7289 & S0(r) & \dots & \dots & \dots & \dots & 21.86 & 33.61$\pm$2.15 & 4.56$\pm$0.63 & trueE \\
NGC7371 & S0/a(r) & 1.10 & 1.46 & 20.00 & 35.49$\pm$2.48 & 21.98 & 35.72$\pm$1.72 & 3.27$\pm$0.37 & bar \\
NGC7377 & S0(s) & 2.62 & 1.63 & 21.09 & 75.53$\pm$19.66 & 20.87 & 52.63$\pm$2.00 & 4.01$\pm$0.33 & \dots \\
NGC7507 & E0 & \dots & \dots & \dots & \dots & 20.08 & 46.97$\pm$1.46 & 4.35$\pm$0.35 & \dots \\
NGC7619 & E2 & \dots & \dots & \dots & \dots & 20.51 & 41.79$\pm$0.78 & 4.34$\pm$0.19 & \dots \\
NGC7778 & E0 & 1.34 & 1.36 & 21.10 & 55.86$\pm$0.29 & 19.75 & 11.40$\pm$0.29 & 3.40$\pm$0.10 & trueS0 \\
NGC7824 & Sab & \dots & \dots & \dots & \dots & 20.70 & 21.32$\pm$1.02 & 4.47$\pm$0.50 & no disk? \\
PGC64863 & E4 & \dots & \dots & \dots & \dots & 21.33 & 19.37$\pm$1.43 & 3.29$\pm$0.50 & inner disk \\
PGC67633 & S0 & \dots & \dots & \dots & \dots & 21.80 & 21.88$\pm$2.05 & 4.32$\pm$0.88 & trueE \\
\enddata
\tablenotetext{a}{Errors shown correspond to 1$\sigma$.}
\end{deluxetable}

\end{document}